\documentclass[nofootinbib]{revtex4}
\usepackage{graphicx}
\usepackage{dcolumn}
\usepackage{amsmath,amssymb,epsfig}
\usepackage{paralist}
\usepackage{comment}
\usepackage{graphicx}

\allowdisplaybreaks

\renewcommand{\vec}[1]{\boldsymbol{\mathrm{#1}}}

\newcommand{\zap}{Zeitschrift f{\"u}r Astrophysik}
\newcommand{\aap}{Astronomy \& Astrophysics}
\newcommand\ddfrac[2]{{\displaystyle\frac{\displaystyle #1}{\displaystyle #2}}}

\begin{document}

\title{Spectrally resolved imaging with the solar gravitational lens}

\author{Slava G. Turyshev$^{1}$, Viktor T. Toth$^2$}

\affiliation{\vskip 3pt$^1$Jet Propulsion Laboratory, California Institute of Technology,\\
4800 Oak Grove Drive, Pasadena, CA 91109-0899, USA}

\affiliation{\vskip 3pt $^2$Ottawa, Ontario K1N 9H5, Canada}

\date{\today}

\begin{abstract}

We consider the optical properties of the solar gravitational lens (SGL) treating the Sun as a massive compact body. Using our previously developed wave-optical treatment of the SGL, we convolve it with a thin-lens representing an optical telescope, and estimate the power spectral density and associated photon flux at individual pixel locations on the image sensor at the focal plane of the telescope. We also consider the solar corona, which is the dominant noise source when imaging faint objects with the SGL. We evaluate the signal-to-noise ratio at individual pixels as a function of wavelength. To block out the solar light, we contrast the use of a conventional internal coronagraph with a Lyot-stop to an external occulter (i.e., starshade). An external occulter, not being a subject to the diffraction limit of the observing telescope, makes it possible to use small telescopes (e.g., $\sim 40$~cm) for spatially and spectrally resolved imaging with the SGL in a broad range of wavelengths from optical to mid-infrared (IR) and without the substantial loss of optical throughput that is characteristic to internal devices. Mid-IR observations are especially interesting as planets are self-luminous at these wavelengths, producing a strong signal, while there is significantly less noise from the solar corona. This part of the spectrum contains numerous features of interest for exobiology and biosignature detection. We develop tools that may be used to estimate instrument requirements and devise optimal observing strategies to use the SGL for high-resolution, spectrally resolved imaging, ultimately improving our ability to confirm and study the presence of life on a distant world.

\end{abstract}


\maketitle

\section{Introduction}
\label{sec:aintro}

Direct obervation of exoplanets is a challenging task as these targets are not self-luminous, they are small, very distant and are moving in a very highly light-contaminated environment \cite{Traub-Oppenheimer:2010,Gaudi:2013}. The thought of resolved imaging of terrestrial exoplanets elevates this problem to the next level, requiring prohibitively large telescopes or interferometric baselines, making the use of conventional astronomical techniques unsuitable for this purpose \cite{Turyshev-Toth:2020-extend,Turyshev-Toth:2022-wil_SNR}.

The solar gravitational lens (SGL) is the only realistic means to overcome these challenges. The SGL exists because the solar gravitational field diffracts electromagnetic (EM) waves that travel in solar proximity \cite{Turyshev:2017,Turyshev-Toth:2017,Turyshev-Toth:2018}. After passing by the Sun, the wavefront converges towards the optical axis, an imaginary line connecting the center of the Sun and that of the source. This region of convergence, the SGL focal region, is at heliocentric distances beyond $R^2_\odot/2r_g\simeq 547.8$ astronomical units (AU), where $R_\odot$ and $r_g$ are the solar geometric and Schwarzschild radii, correspondingly.

A spacecraft, equipped with a modest telescope and an occulter to block sunlight, looking back at the Sun from the SGL's focal region, in proximity to the optical axis corresponding to a distant source such as an exoplanet, will observe the Einstein ring formed around the Sun from amplified light due to that source \cite{Turyshev-Andersson:2002}.

In \cite{Turyshev-Toth:2017,Turyshev-Toth:2018,Turyshev-Toth:2018-grav-shadow} we developed a wave-optical treatment of the SGL by considering diffraction of EM waves in the solar gravity field. In \cite{Turyshev-Toth:2018-plasma,Turyshev-Toth:2019} we considered the propagation of light through the solar corona and found that at infrared (IR), optical and shorter wavelengths, light is practically unaffected by the plasma. We described the image formation process for faint sources \cite{Turyshev-etal:2018,Turyshev-etal:2018-wp,Turyshev-etal:2019-Astro2020-wp,Turyshev-etal:2020-PhaseII} and extended sources \cite{Turyshev-Toth:2019-extend,Turyshev-Toth:2019-blur} at large, but finite distances. As expected, we found that the SGL acts as a lens with significant negative spherical aberration, blurring the images of such extended sources. In \cite{Turyshev-Toth:2019-blur} we studied photometric imaging with the SGL for both point and extended sources and developed analytical expressions for their treatment. We also addressed the imaging of realistic extended sources with an optical telescope that is positioned within the image plane in the SGL strong interference region \cite{Turyshev-Toth:2019-image}. This allowed us to develop robust estimates of the signal-to-noise ratio (SNR) for imaging of realistic exoplanetary sources \cite{Turyshev-Toth:2020-image,Turyshev-Toth:2020-extend,Toth-Turyshev:2020,Turyshev-Toth:2022-wil_SNR}.

As a result of these efforts, we now have all the tools required for the realistic assessment of the SNR that characterize SGL observations of exoplanets, treating them as extended, resolved, faint sources located at large but finite distance from us. In previous studies, we offered broadband estimates, treating the optical telescope as a single sensor. In the present paper, we take the next step and assess the SNR of the spectrally resolved signal on the spatially resolved Einstein ring that is seen to appear around the Sun by the optical telescope. This leads us to an improved understanding of the relationship between SNR and spatial or spectral resolution.

In all previous studies, we assumed that the observing spacecraft is equipped with a telescope carrying a conventional (Lyot-type) internal coronagraph. As it is known, such a coronagraph is subject to the diffraction-limited resolution of the telescope. This precludes the use of modest telescopes with a meter-scale aperture to make SGL observations in the mid-IR band, as such telescopes do not have the resolving power to see the solar disk from the SGL focal region and thus will not be able to use an internal occulter to block solar light. This is unfortunate, since the mid-IR, thermal part of the EM spectrum contains many features of interest that are relevant to lifesign detection.

In the present study, we address an important alternative to the internal coronagraph: An external occulter, known as a starshade, that blocks sunlight while flying in formation with the SGL observing spacecraft. Such a starshade is not subject to the diffraction limit of the observing telescope, and permits the use of smaller telescopes (even telescopes that are too small to resolve the Einstein ring around the Sun!) in conjunction with the SGL.

This paper is organized as follows:
In Section~\ref{sec:signals}, we estimate the anticipated signal while imaging a typical exoplanet. We describe the Fourier-transformed amplitude of the EM field, the intensity distribution and power at the Einstein ring that forms in the focal plane of an optical telescope. We also address the issue of modeling the spectral density of the signals.
In Section~\ref{sec:noise}, we evaluate the stochastic noise contribution of the solar corona.
In Section~\ref{sec:SNR-all} we develop our main result, estimating the SNR in the form of a spectral density, and extend our analysis to mid-IR wavelengths assuming the use of an external starshade.
We present a summary of our results in Section~\ref{sec:concl} and discuss next steps.

 \section{Evaluating the SGL-amplified signals}
\label{sec:signals}

The subject of our investigation is an Earth-like exoplanet, as viewed from the focal region of the SGL, starting at $\sim$548 AU from the Sun, through a thin-lens telescope. Specifically we consider, as the target of observation, an exoplanet in a solar system in our galactic neighborhood, at distances up to $\sim$30 parsec (pc) from the Sun. The image of such an exoplanet is projected by the SGL to an image area several kilometers in size ($\sim$1.3~km for an Earth-like exoplanet at 30~pc, observed at 650~AU from the Sun). A telescope in the focal region of the SGL, looking back in the direction of the Sun, sees a faint Einstein ring form around the Sun from light reflected and emitted by the exoplanet.

We begin with a spectral model of light from the exoplanet, followed by a discussion of the formalism describing image formation with the SGL.

\subsection{Modeling the spectral signal}

\begin{figure}
\includegraphics[]{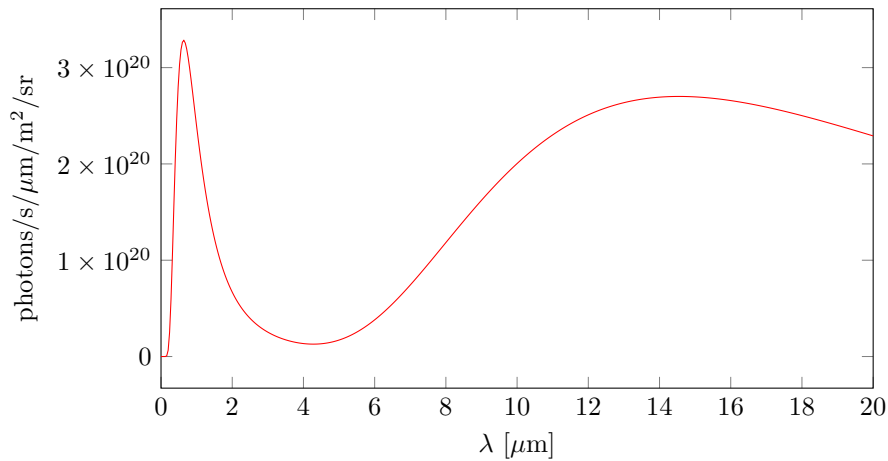}
\caption{\label{fig:Qs}The photon flux from an Earth-like planet, $(\lambda/hc)B_{\tt s}(\lambda)$, measured at the point of emission, according to
(\ref{eq:model-L0IR*}).}
\end{figure}

To provide  estimates for anticipated photon fluxes from realistic exoplanetary targets when they are imaged with the SGL, we model the spectral signal using our own Sun and the Earth as representative cases. Following \cite{Turyshev-Toth:2020-extend}, we consider a planet that is identical to our Earth, orbiting, at a distance of 1 AU, a star that is identical to our Sun. The total flux received by such a target is the same as the solar irradiance at the top of Earth's atmosphere, given as
{}
\begin{align}
I_0=
\sigma  T^4_\odot \Big(\frac{R_\odot}{ {\rm AU}}\Big)^2
&{}=
\pi \Big(\frac{R_\odot}{ {\rm AU}}\Big)^2\int_0^\infty B_\lambda(T_\odot)d\lambda
=\pi \Big(\frac{R_\odot}{ {\rm AU}}\Big)^2\int_0^\infty \frac{2hc^2}{\lambda^5\big(e^{hc/\lambda k_B T_\odot}-1\big)}d\lambda
=1,366.83 ~~   \frac{\rm W}{{\rm m}^2},
\label{eq:model-L0*}
\end{align}
where we use a blackbody radiation model with $\sigma$ as the Stefan-Boltzmann constant, $k_B$ the Boltzmann constant and $T_\odot=5,772$\,K being the temperature of the Sun. Approximating the planet as a Lambertian sphere illuminated from the viewing direction yields a Bond spherical albedo \cite{Turyshev-Toth:2020-extend} of $2/3$, and  the target's average surface brightness becomes
{}
\begin{equation}
B_{\tt s}=\frac{2\alpha}{3\pi} I_0=88.76
~~  \frac{\rm W}{{\rm m}^2\,{\rm sr}},
\label{eq:model-L0B*}
\end{equation}
where we take the Earth's broadband albedo to be $\alpha=0.306$ and assume a fully-illuminated planet at 0 phase angle.

In a more realistic model of the spectral brightness $B_s(\lambda)$ of the exoplanet that includes longer wavelengths, we may also add the planetary thermal emission:
{}
\begin{equation}
B_{\tt s}(\lambda)= \tfrac{2}{3}\alpha
\Big(\frac{R_\odot}{ {\rm AU}}\Big)^2 \frac{2hc^2}{\lambda^5\big(e^{hc/\lambda k_B T_\odot}-1\big)}+ \frac{2hc^2}{\lambda^5\big(e^{hc/\lambda k_B T_\oplus}-1\big)} ~~ \frac{\rm W}{\mu{\rm m}\,{\rm m}^2\,{\rm sr}},
\label{eq:model-L0IR*}
\end{equation}
where $T_\oplus=252$\,K is the effective radiating temperature of the Earth\footnote{See \url{https://en.wikipedia.org/wiki/Effective_temperature}}. The resulting photon flux is found, as usual, as $Q_{\tt s}(\lambda)=(\lambda/hc)B_{\tt s}(\lambda)$ and is shown in (Fig.~\ref{fig:Qs}). Integrated over all wavelengths, we obtain
\begin{align}
Q_s=\int_0^\infty d\lambda \,\frac{\lambda}{hc}B_{\tt s}(\lambda)=
\frac{4 k^3\zeta(3)}{c^2h^3}\left(\tfrac{2}{3}\alpha\Big(\frac{R_\odot}{ {\rm AU}}\Big)^2   T_\odot^3+T_\oplus^3\right)=4.1\times 10^{20}
+7.75\times 10^{21} ~~~   \frac{\rm photons}{{\rm s}\, {\rm m}^2\,{\rm sr}},
\end{align}
where $\zeta(3)\sim 1.202$ is the Riemann zeta function. The first term represents mostly photons in the optical and near IR part of the spectrum, whereas the second term corresponds to the much greater number of thermal IR photons.

\subsection{Image formation process with the SGL}
\label{sec:im-form}

Following \cite{Turyshev-Toth:2020-extend}, we consider a light ray with the wavevector $\vec k=(0,0,1)$. To discuss imaging, we consider the source plane, image plane, and the optical telescope's focal (or sensor) plane, each of them being orthogonal to ${\vec k}$. Next, we introduce two-dimensional coordinates to describe points in the source plane, $\vec x'$; the position of the telescope in the image plane,  $\vec x_0$; points in the image plane within the telescope's aperture, $\vec x$; and points in the optical telescope's focal plane ${\vec x}_i$. These are given as  follows:
\begin{eqnarray}
\{{\vec x}'\}&\equiv& (x',y')=\rho'\big(\cos\phi',\sin\phi'\big)=\rho'{\vec n}',
\label{eq:x'}\\
\{{\vec x}_0\}&\equiv& (x_0,y_0)=\rho_0\big(\cos\phi_0,\sin\phi_0\big)=\rho_0{\vec n}_0, \label{eq:x0}\\
\{{\vec x}\}&\equiv& (x,y)=\rho\big(\cos\phi,\sin\phi\big)=\rho\,{\vec n}, \label{eq:x}\\
 \{{\vec x}_i\}&\equiv& (x_i,y_i)=\rho_i\big(\cos\phi_i,\sin\phi_i\big)=\rho_i{\vec n}_i.
  \label{eq:p}
\end{eqnarray}

We rely on  (\ref{eq:x'})--(\ref{eq:p}), but slightly redefining them  by introducing ${\vec x}_0=-({\overline z}/{z_0}){\vec x}_0'
$ that allows us to introduce
{}
\begin{eqnarray}
{\vec x}''={\vec x}'-{\vec x}'_0\equiv \rho''{\vec n}''= \rho''(\cos\phi'',\sin\phi'').
  \label{eq:coord2}
\end{eqnarray}

We use the following notations for the two spatial frequencies $\alpha$ and $\eta_i$, and a useful ratio $\beta$:
{}
\begin{eqnarray}
\alpha&=&k \sqrt{\frac{2r_g}{\overline z}}, \qquad \eta_i=k\frac{\rho_i}{f}, \qquad \beta=\frac{\overline z}{{z}_0},
  \label{eq:alpha-mu}\\
 \vec \alpha&=&(\alpha_x,\alpha_y)=\alpha \big(\cos\phi_\xi,\sin\phi_\xi\big)=\alpha \vec n_\xi, \qquad \vec \eta_i=\eta_i \vec n_i.
 \label{eq:vec_alpha-eta}
\end{eqnarray}

With these definitions, the intensity distribution on the image sensor of a telescope corresponding to a signal received from an exoplanet is given as
  {}
\begin{eqnarray}
I_{\tt }({\vec x}_i,{\vec x}_0) &=&
\frac{1}{z^2_0}\mu_0 \Big(\frac{kd^2}{8f}\Big)^2\hskip-4pt
\iint d^2{\vec x}''  B_{\tt s}(\lambda)  {\cal A}^2({\vec x}_i,{\vec x}''), \qquad \mu_0=2\pi kr_g,
  \label{eq:pow-blur}
\end{eqnarray}
where in the case of a monopole SGL, the Fourier-transformed amplitude of the EM field ${\cal A}({\vec x}_i,{\vec x}'')$ has the form:
{}
\begin{eqnarray}
{\cal A}({\vec x}_i,{\vec x}'')&=&
 \frac{1}{2\pi}\int_0^{2\pi} d\phi_\xi \Big(\frac{2J_1(\alpha d \, \hat u(\phi_\xi,\vec x_i))}{\alpha d \, \hat u(\phi_\xi,\vec x_i)}\Big)\exp\Big[-i \alpha\beta \rho''\cos(\phi_\xi-\phi'')\Big],
  \label{eq:amp-mono}
\end{eqnarray}
where $\alpha$ and $\beta$ are from (\ref{eq:alpha-mu}) and the normalized spatial frequency $\hat u(\phi_\xi,\vec x_i)$ has the form (see \cite{Turyshev-Toth:2020-extend})
{}
\begin{eqnarray}
\hat u(\phi_\xi,\vec x_i)=
\big| \vec \alpha + \vec \eta_i\big|/2\alpha&=&
\Big\{{\textstyle\frac{1}{4}}\Big(1-\frac{\eta_i}{\alpha}\Big)^2+\frac{\eta_i}{\alpha}\cos^2[{\textstyle\frac{1}{2}}\big(\phi_\xi-\phi_i\big)] \Big\}^\frac{1}{2}.
\label{eq:upm}
\end{eqnarray}

The integral (\ref{eq:pow-blur}) must be evaluated for two different regions corresponding to the telescope pointing within the image and outside of it, as was done in \cite{Turyshev-Toth:2019-blur}. The principal technical challenge is the evaluation of the integral
(\ref{eq:amp-mono}) that represents a Fourier-transform of the EM field amplitude.

Equation~(\ref{eq:amp-mono}) describes a monopole gravitational lens. The SGL, of course, has a small but non-trivial quadrupole mass moment (see details in \cite{Turyshev-Toth:2021-multipoles,Turyshev-Toth:2021-caustics,Turyshev-Toth:2021-quartic}). While the resulting change in its optical properties has significant consequences when it comes to image reconstruction through deconvolution, the amount of light from an extended source, such as an exoplanet, deposited in the SGL image plane remains approximately the same until and unless the source's projected image becomes small enough to be comparable in size to the caustic pattern projected by the quadrupole moment. The actual size of the caustic pattern depends on solar latitude, but even in the solar equatorial plane, it remains a small fraction of the projected size of an Earth-like exoplanet at $z_0\lesssim 30$~pc. Therefore, the signal flux remains unaffected, and it is safe to use the monopole Sun to estimate signal and noise levels.

\subsection{Fourier-transformed amplitude of the EM field}

The integral (\ref{eq:amp-mono}) can be evaluated using the method of stationary phase. With the rapidly varying phase given as
{}
\begin{eqnarray}
\varphi(\phi_\xi)=-\alpha\beta \rho''\cos(\phi_\xi-\phi''),
  \label{eq:ph-mono}
\end{eqnarray}
we compute the first and second derivatives of this expression:
{}
\begin{eqnarray}
\varphi'(\phi_\xi)&=&\alpha\beta \rho''\sin(\phi_\xi-\phi'')
\qquad {\rm and} \qquad
\varphi''(\phi_\xi)=\alpha\beta \rho''\cos(\phi_\xi-\phi'').
  \label{eq:ph-mono-der}
\end{eqnarray}
The phase is stationary when $\varphi'(\phi_\xi)=\alpha\beta \rho''\sin(\phi_\xi-\phi'')=0$. Thus, we have two solutions:
{}
\begin{eqnarray}
\phi_\xi-\phi''=\{0,\pi\}.
  \label{eq:ph-mono-sol}
\end{eqnarray}
With these solutions, we compute the corresponding expressions for the phase and its second derivative:
{}
\begin{eqnarray}
\big\{\varphi_0,\varphi''_0\big\}&=&\big\{-\alpha\beta \rho'',\alpha\beta \rho''\big\} \qquad {\rm and} \qquad
\big\{\varphi_\pi,\varphi''_\pi\big\}=\big\{\alpha\beta \rho'',-\alpha\beta \rho''\big\}.
  \label{eq:ph-mono-sol2}
\end{eqnarray}

As a result, the integral (\ref{eq:amp-mono}) takes the form
{}
\begin{eqnarray}
{\cal A}({\vec x}_i,{\vec x}'')&=&
 \frac{1}{\sqrt{2\pi \alpha\beta\rho''}}\Big\{\Big(\frac{2J_1(\alpha d \, \hat u_0(\phi'',\vec x_i))}{\alpha d \, \hat u_0(\phi'',\vec x_i)}\Big)e^{-i \big(\alpha\beta \rho''-\frac{\pi}{4}\big)}+
 \Big(\frac{2J_1(\alpha d \, \hat u_\pi(\phi'',\vec x_i))}{\alpha d \, \hat u_\pi(\phi'',\vec x_i)}\Big) e^{i \big(\alpha\beta \rho''-\frac{\pi}{4}\big)}\Big\},
  \label{eq:amp-mono-int}
\end{eqnarray}
and its square, ${\cal A}^2({\vec x}_i,{\vec x}'')$ that is needed in (\ref{eq:pow-blur}), is given as
{}
\begin{eqnarray}
{\cal A}^2({\vec x}_i,{\vec x}'')&=&
 \frac{1}{2\pi \alpha\beta\rho''}\Big\{\Big(\frac{2J_1(\alpha d \, \hat u_0(\phi'',\vec x_i))}{\alpha d \, \hat u_0(\phi'',\vec x_i)}\Big)^2+
 \Big(\frac{2J_1(\alpha d \, \hat u_\pi(\phi'',\vec x_i))}{\alpha d \, \hat u_\pi(\phi'',\vec x_i)}\Big)^2+\nonumber\\
 &&\hskip 50pt +~ 2\sin(2\alpha\beta\rho'')\Big(\frac{2J_1(\alpha d \, \hat u_0(\phi'',\vec x_i))}{\alpha d \, \hat u_0(\phi'',\vec x_i)}\Big)\Big(\frac{2J_1(\alpha d \, \hat u_\pi(\phi'',\vec x_i))}{\alpha d \, \hat u_\pi(\phi'',\vec x_i)}\Big)\Big\},
  \label{eq:amp-mono-int2}
\end{eqnarray}
where $\hat u_0(\phi'',\vec x_i)$ and $\hat u_\pi(\phi'',\vec x_i)$ from (\ref{eq:upm}) with solutions for $\phi_\xi$ from (\ref{eq:ph-mono-sol}) are given as
{}
\begin{eqnarray}
\hat u_0(\phi_\xi,\vec x_i)&=&
\Big\{{\textstyle\frac{1}{4}}\Big(1-\frac{\eta_i}{\alpha}\Big)^2+\frac{\eta_i}{\alpha}\cos^2[{\textstyle\frac{1}{2}}\big(\phi''-\phi_i\big)] \Big\}^\frac{1}{2},
 \qquad
\hat  u_\pi(\phi_\xi,\vec x_i)=
\Big\{{\textstyle\frac{1}{4}}\Big(1-\frac{\eta_i}{\alpha}\Big)^2+\frac{\eta_i}{\alpha}\sin^2[{\textstyle\frac{1}{2}}\big(\phi''-\phi_i\big)] \Big\}^\frac{1}{2}.~~~~
  \label{eq:upm-n0+}
\end{eqnarray}

Using the result (\ref{eq:upm-n0+}), we recognize that at the Einstein ring, where ${\eta_i}=\alpha$, the spatial frequencies become $\hat u_0(\phi_\xi,\vec x_i)=\cos[{\textstyle\frac{1}{2}}\big(\phi''-\phi_i\big)] $ and $\hat  u_\pi(\phi_\xi,\vec x_i)=
\sin[{\textstyle\frac{1}{2}}\big(\phi''-\phi_i\big)]$, implying that for any given $\phi_i$ there always will be $\phi''$ such that ${\textstyle\frac{1}{2}}\big(\phi''-\phi_i\big)$ is either $\pm {\textstyle\frac{1}{2}}\pi$ or $\pm\pi$. At these points, either $\hat u_0(\phi_\xi,\vec x_i)=0$ or $\hat u_\pi(\phi_\xi,\vec x_i)=0$, so that the ratios of the Bessel-functions in (\ref{eq:amp-mono-int2}) reach their largest value $2J_1(x)/x \rightarrow 1$. As a result, (\ref{eq:amp-mono-int2}) yields the well-known expression for the point-spread function (PSF) of the monopole lens (e.g., see \cite{Turyshev-Toth:2020-extend} an references therein):
{}
\begin{eqnarray}
{\rm PSF}({\vec x}_i,{\vec x}'')&=&
  \frac{1}{\pi \alpha\beta\rho''}\Big(1+\sin2\alpha\beta\rho''\Big)\equiv  \frac{2}{\pi \alpha\beta\rho''}\cos^2\Big(\alpha\beta\rho''-{\textstyle\frac{\pi}{4}}\Big),
  \label{eq:psf-mono}
\end{eqnarray}
which represents an approximation of $J^2_0(\alpha\beta\rho'')$, as it should be for the monopole gravitational lens.

\subsection{Spectral intensity on the image sensor}
\label{sec:intense-in-m}

Expression (\ref{eq:pow-blur}) allows us to compute the power received from the resolved source.  To do that, we introduce a new coordinate system in the source plane, ${\vec x}''$, with the origin at the center of the directly imaged region: ${\vec x}'-{\vec x}'_0={\vec x}''$. As vector ${\vec x}'_0$ is constant, $dx'dy'=dx''dy''$. Next, in the new coordinate system, we use polar coordinates $(x'',y'')\rightarrow (r'',\phi'')$. In these coordinates, the circular edge of the source, $R_\oplus$, is no longer a circle but a curve, $\rho_\oplus(\phi'')$, the radial distance of which is given by the following relation:
{}
\begin{eqnarray}
\rho_\oplus(\phi'')
&=&\sqrt{R_\oplus^2-{\rho'_0}^2\sin^2\phi''}-\rho'_0\cos\phi''.
\label{eq:rho+}
\end{eqnarray}

For an actual astrophysical source, $B_s({\vec x}',\lambda)$ is, of course, an arbitrary function of the coordinates ${\vec x}'$ and thus the integral (\ref{eq:pow-blur}) can only be evaluated numerically. However, we can obtain an analytic result in the simple case of a disk of uniform brightness, characterized by ${B}_s({\vec x}',\lambda) ={B}_s(\lambda)$. In this case, we integrate (\ref{eq:pow-blur}):
  {}
\begin{eqnarray}
I_{\tt }({\vec x}_i,{\vec x}_0,\lambda) &=&
 \mu_0\frac{B_{\tt s}(\lambda)}{z^2_0} \Big(\frac{kd^2}{8f}\Big)^2  \int_0^{2\pi} \hskip -8pt d\phi'' \int_{0}^{\rho_\oplus}\hskip -4pt \rho''d\rho''  {\cal A}^2({\vec x}_i,{\vec x}'').
  \label{eq:intense}
\end{eqnarray}

Considering only the monopole solar gravitational field, we use (\ref{eq:amp-mono-int2}), and rewrite (\ref{eq:intense}) as
{}
\begin{eqnarray}
I_{\tt }({\vec x}_i,{\vec x}_0,\lambda)  &=&
\frac{B_{\tt s}(\lambda)}{z^2_0} \Big(\frac{kd^2}{8f}\Big)^2    \frac{ \mu_0R_\oplus}{\alpha\beta} \times\nonumber\\
&&\hskip -55pt \times\,
\frac{ 1}{2\pi} \int_0^{2\pi} \hskip -8pt d\phi'' \bigg\{
\Big(\sqrt{1-\Big(\frac{\rho_0}{r_\oplus}\Big)^2\sin^2\phi''}-\frac{\rho_0}{r_\oplus}\cos\phi''\Big)
\bigg(\Big(\frac{2J_1(\alpha d\, \hat u_0(\phi'',\vec x_i))}{\alpha d\, \hat u_0(\phi'',\vec x_i)}\Big)^2+
 \Big(\frac{2J_1(\alpha d\, \hat u_\pi(\phi'',\vec x_i))}{\alpha d\, \hat u_\pi(\phi'',\vec x_i)}\Big)^2\bigg)+\nonumber\\
 &&\hskip -40pt +~
\frac{ 2}{\alpha r_\oplus}\sin^2\Big[\alpha r_\oplus
\Big(\sqrt{1-\Big(\frac{\rho_0}{r_\oplus}\Big)^2\sin^2\phi''}-\frac{\rho_0}{r_\oplus}\cos\phi''\Big)\Big]
\Big(\frac{2J_1(\alpha d\, \hat u_0(\phi'',\vec x_i))}{\alpha d\, \hat u_0(\phi'',\vec x_i)}\Big)\Big(\frac{2J_1(\alpha d\, \hat u_\pi(\phi'',\vec x_i))}{\alpha d\, \hat u_\pi(\phi'',\vec x_i)}\Big)\bigg\},
  \label{eq:intense1*}
\end{eqnarray}
where
$\hat u_0(\phi'',\vec x_i)$ and $\hat u_\pi(\phi'',\vec x_i)$ introducted by (\ref{eq:upm-n0+})  with the help of definitions (\ref{eq:alpha-mu})--(\ref{eq:vec_alpha-eta}) are given as below
{}
\begin{eqnarray}
\hat u_0(\phi'',\vec x_i)&=&
\Big\{{\textstyle\frac{1}{4}}\Big(1-\frac{\rho_i}{f}\Big(\frac{\overline z}{2r_g}\Big)^\frac{1}{2}\Big)^2+\frac{\rho_i}{f}\Big(\frac{\overline z}{2r_g}\Big)^\frac{1}{2}\cos^2[{\textstyle\frac{1}{2}}\big(\phi''-\phi_i\big)] \Big\}^\frac{1}{2},
  \label{eq:mono-u-hat-0+} \\
  \hat u_\pi(\phi'',\vec x_i)&=&
\Big\{{\textstyle\frac{1}{4}}\Big(1-\frac{\rho_i}{f}\Big(\frac{\overline z}{2r_g}\Big)^\frac{1}{2}\Big)^2+\frac{\rho_i}{f} \Big(\frac{\overline z}{2r_g}\Big)^\frac{1}{2}\sin^2[{\textstyle\frac{1}{2}}\big(\phi''-\phi_i\big)]\Big\}^\frac{1}{2},
  \label{eq:mono-u-hat-pi+}
\end{eqnarray}
both independent on the wavelength. The expression $({\rho_i}/{f})/ \sqrt{{2r_g}/{\overline z}}$ is the ratio of the angle corresponding to a particular pixel on the image sensor $\theta_i={\rho_i}/{f}$ to the angle corresponding to the Einstein ring $\theta_{\tt ER}=\sqrt{{2r_g}/{\overline z}}$.

We observe that the ratios involving the Bessel-functions in the expression (\ref{eq:intense1*}) above are at most $2J_1(x)/x=1$, at $x=0$. Given the fact that the spatial frequency $\alpha$ is quite high, for most values of the argument these ratios become negligible. We also observe that the last term in this expression is at most $\propto 2/\alpha r_\oplus \ll1$, which is negligibly small compared to the first term in this expression. Thus, the last term in (\ref{eq:intense1*}) can be omitted with this expression taking the form (see details in \cite{Turyshev-Toth:2020-extend})
  {}
\begin{eqnarray}
I_{\tt }({\vec x}_i,{\vec x}_0,\lambda)  &\simeq&
\frac{B_{\tt s}(\lambda)}{z^2_0} \Big(\frac{kd^2}{8f}\Big)^2    \frac{ \mu_0R_\oplus}{\alpha\beta}
\times\nonumber\\
&& \times\,
\frac{ 1}{2\pi} \int_0^{2\pi} \hskip -8pt d\phi''
\sqrt{1-\Big(\frac{\rho_0}{r_\oplus}\Big)^2\sin^2\phi''}
\bigg[\Big(\frac{2J_1(\alpha d \,\hat u_0(\phi'',\vec x_i))}{\alpha d \,\hat u_0(\phi'',\vec x_i)}\Big)^2+
 \Big(\frac{2J_1(\alpha d \,\hat u_\pi(\phi'',\vec x_i))}{\alpha d \,\hat u_\pi(\phi'',\vec x_i)}\Big)^2\bigg],~~~~
  \label{eq:mono-intense2*}
\end{eqnarray}
where we obtained the final form of the equation by dropping the $(\rho_0/r_\oplus)\cos\phi''$ term in the first integral in (\ref{eq:mono-intense2*}), as this term, multiplied by the squared Bessel-function terms that have the same periodicity by virtue of the dependence of $\hat u_0,\hat u_\pi$ on $\phi''$, vanishes identically when integrated over a full $2\pi$ period.

The photon count density (per unit time, unit wavelength, and unit area) that corresponds to (\ref{eq:mono-intense2*}) can be readily calculated:
\begin{align}
Q_{\tt }({\vec x}_i,{\vec x}_0,\lambda)=\frac{\lambda}{hc}I_{\tt }({\vec x}_i,{\vec x}_0,\lambda).
\label{eq:Qsignal}
\end{align}
This quantity is of primary interest as it forms the basis for calculating stochastic shot noise, which results from the quantized nature of light.

\subsection{Evaluating the spectral signals}

\begin{figure}
\includegraphics[]{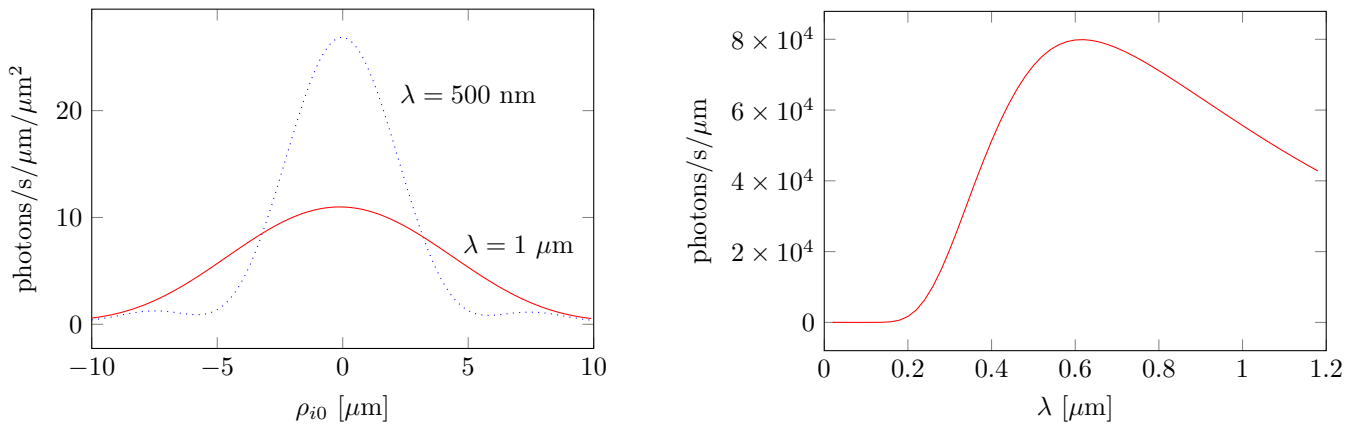}
\caption{\label{fig:Qfppix-c}Left: Spectral flux of photons per square micron sensor pixel, at $\lambda=500~$nm (blue dotted line) and $\lambda=1~\mu$m (red solid line). We present a cross-sectional view, $\rho_{i0}=\rho_i-f\sqrt{2r_g/{\overline z}}$, centered on the Einstein ring that is due to an Earth-like planet at a distance of $z_0=30$~pc, as seen on the image sensor of an $f=10$~m thin-lens telescope with aperture $d=1$~m, placed at $\rho_0=0$ at an image plane at ${\overline z}=650$~AU from the Sun, in accordance with Eqs.~(\ref{eq:Qsignal}) and (\ref{eq:model-L0IR*}). Right: Photon spectral density integrated over the entire Einstein ring. Note that this computation assumes a resolved Einstein ring, which implies $\lambda\lesssim 1.2~\mu$m for a $d=1$~m telescope.}
\end{figure}

The aperture of an optical telescope determines its diffraction limit and, ultimately, its ability to resolve the Einstein ring and distinguish it from the Sun at various wavelengths. This is especially important when a conventional internal coronagraph is used in conjunction with a Lyot-stop \cite{Turyshev-Toth:2020-extend} (see the discussion in Sec.~\ref{sec:sol-coronagraph-model}). For a $d=1$~m telescope, this represents a realistic limit of $\lambda\sim 1.2~\mu$m before the telescope can no longer reliably distinguish the solar disk (and thus make it possible for an internal occulter to block its light) and the Einstein ring, viewed from a distance of ${\overline z}=650$~AU. Therefore, we consider such an instrument only for use in optical and near-IR wavelengths.

Two representative cases are depicted in Fig.~\ref{fig:Qfppix-c}~(left), showing the photon spectral density at two different wavelengths, $\lambda=500$~nm and $\lambda=1~\mu$m, in the form of a cross-sectional view of the Einstein ring, as it appears on the image sensor in a $f=10$~m thin-lens telescope, according to (\ref{eq:mono-intense2*}). As expected, at shorter wavelengths the Einstein ring is much sharper, and the peak photon count is higher despite the fact that these are higher energy photons and therefore, there are fewer of them for a given light intensity.

Nonetheless, the photon counts are very low for the target considered, an exo-Earth at $z_0=30$~pc. As we shall see in the next section, the photon count due to the solar corona is much higher, so we are looking at a faint signal on a bright background. It is for this reason that in all previous analyses, we considered integrating the signal over the entire Einstein ring, using the telescope as a ``light bucket'', a single-pixel sensor that traverses the (kilometer-scale) SGL image plane, sampling it one pixel at a time. Integrating the signal over the entire Einstein ring yields the spectral density shown in Fig.~\ref{fig:Qfppix-c}~(right).

To estimate the photon count  a broadband spatially resolved image, we need to integrate not just over the Einstein ring but over all wavelengths. Numerical integration up to $\lambda=1.2~\mu$m yields a photon count of $\sim 5.59\times 10^4$ photons per second. This figure is lower than the $\sim 8\times 10^4$ photons we estimated in previous work; the reason for this difference is that in these previous analyses, we used $\lambda=1~\mu$m as the characteristic wavelength of the broadband signal, whereas the actual peak of the solar blackbody is near $\lambda=500$~nm. This skews the result towards fewer photons of higher energy.

\section{Noise from the solar corona}
\label{sec:noise}

In order to evaluate the feasibility of using the SGL for imaging and to assess the required light collection (integration) times and achievable resolution, it is essential to have a clear and accurate understanding of the amount of noise that is present in SGL observations.
This is especially important when we consider the possibility of reconstructing high resolution images of a distant source. Whether the reconstruction uses details of an observed Einstein ring, observations of varying light intensity in the SGL image plane, or a combination of techniques, reconstruction amounts to inverting the mapping of source light into the image plane by the SGL PSF or the combined PSF of the SGL and the observing instrument. This mapping is encoded in the convolution matrix of the lens (or lens plus instrument). The process of recovering the original image, deconvolution, is known to amplify noise disproportionately at the expense of signal. Successful deconvolution, therefore, requires data to be collected with the highest SNR possible. The limits on the SNR---limits due to keeping integration times reasonable, due to limitations of the image sensor dynamic range, or other factors---therefore represent the primary constraint on the achievable spatial and spectral resolution when imaging distant exoplanets using the SGL.

The Einstein ring, formed by light from the exoplanet and amplified by the SGL, appears near the solar disk, on the bright solar corona background. The brightness of the solar corona dominates over the Einstein ring formed by the SGL from the faint light of distant, dim objects. Noise due to the solar corona will always be present due to the quantized nature of light, in the form of stochastic shot noise. Even assuming that no systematic noise sources are present and that the contribution of the solar corona can be independently measured and removed from any observation, this noise remains, and it represents the main limitation on achievable imaging with the SGL.

\subsection{Solar corona model}

To model the contribution from the solar corona, we use a recent model \cite{November:1996} of its surface brightness $B_{\tt cor}(\theta)$:
{}
\begin{equation}
B_{\tt cor}(\theta)= \Big(10^{-6}\cdot B_\odot\Big)\Big[3.670 \Big(\frac{\theta_0}{\theta}\Big)^{18}+1.939\Big(\frac{\theta_0}{\theta}\Big)^{7.8}+ 5.51\times 10^{-2} \Big(\frac{\theta_0}{\theta}\Big)^{2.5}\Big],
\label{eq:model-th*}
\end{equation}
where $\theta=\rho/{\overline z}$ and $\theta_0=R_\odot/{\overline z}$. The quantity $B_\odot$ is the brightness at the center of the solar disk. Assuming the Sun's temperature\footnote{See \url{https://en.wikipedia.org/wiki/Sun}} to be $T_\odot=5,772$~K, we estimate the solar brightness from Planck's radiation law:
{}
\begin{equation}
B_\odot= {\textstyle\frac{1}{\pi}}\sigma  T^4_\odot=\int_0^\infty B_\lambda(T_\odot)d\lambda =\int_0^\infty \frac{2hc^2}{\lambda^5\big(e^{hc/\lambda k T_\odot}-1\big)}d\lambda=2.0034 \times 10^7 ~~   \frac{\rm W}{{\rm m}^2\,{\rm sr}}.
\label{eq:model-L02*}
\end{equation}

Substituting this result  in (\ref{eq:model-th*}) yields the known broadband corona model \cite{Turyshev-Toth:2020-extend,Turyshev-Toth:2022-wil_SNR},
{}
\begin{equation}
B_{\tt cor}(\theta)= 20.03\Big[3.670 \Big(\frac{\theta_0}{\theta}\Big)^{18}+1.939\Big(\frac{\theta_0}{\theta}\Big)^{7.8}+ 5.51\times 10^{-2} \Big(\frac{\theta_0}{\theta}\Big)^{2.5}\Big]  ~~   \frac{\rm W}{{\rm m}^2\,{\rm sr}}.
\label{eq:model-th}
\end{equation}

\begin{figure}
\includegraphics[]{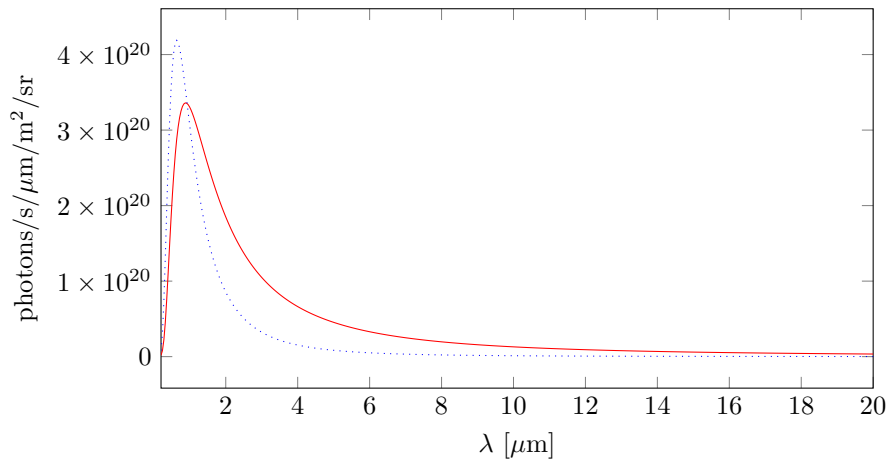}
\caption{\label{fig:Qcor}The modeled photon spectral flux of the solar corona (solid red line), $(\lambda/hc)B_{\tt cor}(\theta,\lambda)$ according to Eq.~(\ref{eq:model-th-lam*}). For comparison, a Planckian spectrum (dotted blue line) is also shown, yielding the same integrated broadband intensity as the model we use.}
\end{figure}

Considering the spectral behavior of the solar corona, we realize that it is not Planckian. Relevant data are scarce, especially in the infrared domain \cite{Boe:2021,DelZanna:2018,Penn:2014,MacQueen:1995,Mann:1992,Rao:1981,Minnaert:1930}. Nonetheless, the few data points from available sources suggest a reasonable approximation in the form of the Planckian spectrum scaled by the wavelength $\lambda$:
\begin{equation}
B_{\tt cor}(\theta,\lambda)= \Big(10^{-6}\frac{\lambda}{\lambda^\star} B_\odot(\lambda)\Big)\Big[3.670 \Big(\frac{\theta_0}{\theta}\Big)^{18}+1.939\Big(\frac{\theta_0}{\theta}\Big)^{7.8}+ 5.51\times 10^{-2} \Big(\frac{\theta_0}{\theta}\Big)^{2.5}\Big],
\label{eq:model-th*-scaled}
\end{equation}
where we use $\lambda^\star= 925$~nm, ensuring that $\int_0^\infty d\lambda B_{\tt cor}(\theta,\lambda)=B_{\tt cor}(\theta)$ is consistent with (\ref{eq:model-th}). Consequently, we model the spectral corona brightness as (see Fig.~\ref{fig:Qcor}):
{}
\begin{eqnarray}
B_{\tt cor}(\theta,\lambda)&=&
10^{-12}
\frac{2hc^2}{\lambda^\star\lambda^4\big(e^{hc/\lambda k T_\odot}-1\big)}\Big[3.670 \Big(\frac{\theta_0}{\theta}\Big)^{18}+1.939\Big(\frac{\theta_0}{\theta}\Big)^{7.8}+ 5.51\times 10^{-2} \Big(\frac{\theta_0}{\theta}\Big)^{2.5}\Big] ~~   \frac{\rm W}{\mu{\rm m}\,{\rm m}^2\,{\rm sr}},
\label{eq:model-th-lam*}
\end{eqnarray}
where another factor of $10^{-6}$ is because we use $\mu$m to report the scale of a wavelength.

The modified spectrum of (\ref{eq:model-th-lam*}) represents a conservative model for estimating the SGL SNR. Skewing the model in favor of the infrared part of the spectrum yields a greater number of photons, hence more shot noise. Therefore, until and unless a better spectral model for the corona becomes available, we find this model satisfactory for the purpose of estimating the SNR of SGL observations.

\subsection{Spectral signal from the solar corona}

\begin{figure}
\includegraphics[]{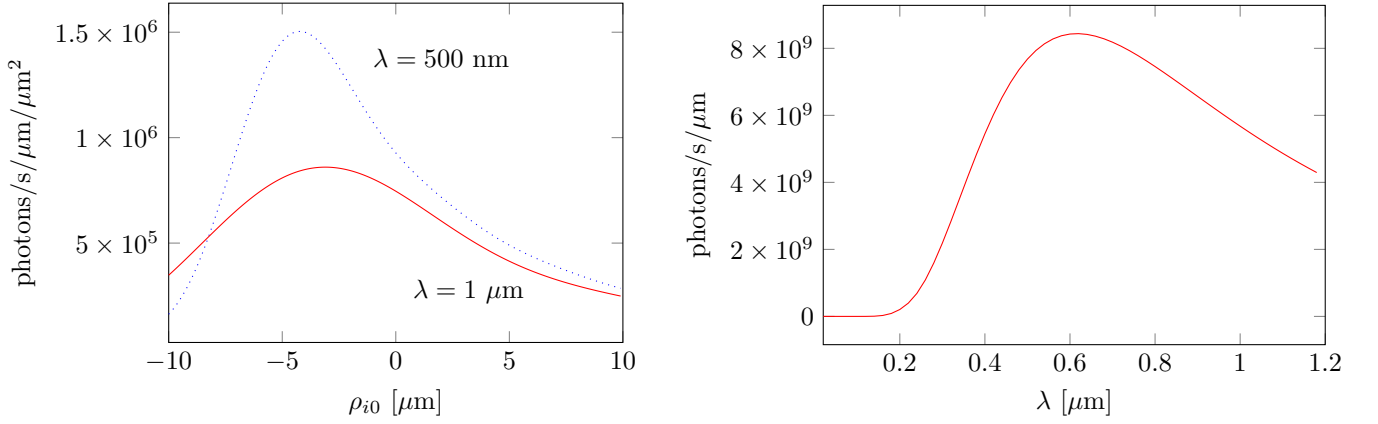}
\caption{\label{fig:Qcorpix-c}Left: Cross-sectional view of the corona photon flux $Q_{\tt cor}$ given by Eq.~(\ref{eq:Qcorona}), centered on the Einstein-ring, $\rho_{i0}=\rho_i-f\sqrt{2r_g/{\overline z}}$, as seen by a $d=1$~m aperture telescope at ${\overline z}=650$~AU from the Sun, in the vicinity of the Einstein-ring of a distant target, at two different wavelengths.
Right: Spectral density of the photon count due to the solar corona, integrated over the Einstein ring. Similarly to Fig.~\ref{fig:Qfppix-c}, the computation is limited to $\lambda<1.2~\mu$m, consistent with diffraction limits.}
\end{figure}

As we established, the Einstein ring that forms around the Sun from light emitted by the exoplanet that is the observational target appears on the bright background of the solar corona. Even if we assume that the corona background can be accurately estimated (or measured by other instruments) and removed, as light is quantized into photons, inevitably, there is stochastic noise in the form of Poisson (approximately Gaussian) shot noise.

The spectral intensity distribution due to corona light received on the sensor of an imaging telescope is evaluated similarly to (\ref{eq:pow-blur}):
  {}
\begin{eqnarray}
I_{\tt cor}({\vec x}_i,\lambda) =
\Big(\frac{kd^2}{8f}\Big)^2
\int_0^{2\pi}\hskip -4pt d\phi'\int_{\theta_{\tt 0}}^{\infty}\hskip -4pt \theta' d\theta'  \, B_{\tt cor}\big(\theta',\lambda\big) \Big( \frac{2
J_1\big(kd\, \hat u(\vec x',\vec x_i)\big)}{kd\, \hat u(\vec x',\vec x_i)}\Big)^2,
  \label{eq:pow-cor*}
\end{eqnarray}
where $\theta_0=R_\odot/{\overline z}$ represents the solar disk, light from which is assumed to be blocked by an internal coronagraph or external starshade; and where, following \cite{Turyshev-Toth:2019-extend}, we introduce the corona spatial frequency, $\vec \alpha_c$ defined as
{}
\begin{eqnarray}
\vec \alpha_c=k\frac{\vec x'}{\overline z}\equiv k\frac{\rho'}{\overline z}\,\vec n'=k\theta'\,\vec n',
  \label{eq:upmAc}
\end{eqnarray}
and using (\ref{eq:alpha-mu}) to represent $\vec \eta_i=\eta_i\,\vec n_i$, we define $\hat u(\vec x',\vec x_i)$, a wavelength-independent corona spatial frequency, as
{}
\begin{eqnarray}
\hat u(\vec x',\vec x_i)=|\vec \alpha_c+\vec \eta_i|/2k=
 \Big\{{\textstyle\frac{1}{4}}\Big(\theta' -\frac{\rho_i}{f}\Big)^2+\theta'\frac{\rho_i}{f}\cos^2[{\textstyle\frac{1}{2}}(\phi'-\phi_i)]\Big\}^\frac{1}{2}.
  \label{eq:upmA}
\end{eqnarray}

Considering the expression for $\hat u(\vec x',\vec x_i)$ given by (\ref{eq:upmA}) and taking into account the integration limits for $\theta'$, one can see that for any pixel with radial position of $\rho_i\geq f \theta_0=f \, R_\odot/\overline z$, there will be an area of the corona with angular position of $\theta' $ such that $\theta' ={\rho_i}/{f}$, yielding  $\hat u(\vec x',\vec x_i)=\theta'\cos[{\textstyle\frac{1}{2}}(\phi'-\phi_i)]$. As a result, for the value of the argument $(\phi'-\phi_i)=\pm\pi$, the ratio of the Bessel-function in (\ref{eq:pow-cor*}) will reach its maximum value of 1. This will result in the light present everywhere on the sensor for pixels with  $\rho_i\geq f \theta_0$. Such a behavior is different compared to that of the exoplanetary signal that is concentrated in the immediate vicinity of the Einstein ring, as observed in (\ref{eq:mono-u-hat-0+}) and (\ref{eq:mono-u-hat-pi+}).

The photon count density (photons per unit time, per unit wavelength, per unit area) that corresponds to (\ref{eq:pow-cor*}),
is shown in Fig.~\ref{fig:Qcorpix-c} (left) and it is given by:
\begin{align}
Q_{\tt cor}({\vec x}_i,\lambda)=\frac{\lambda}{hc}I_{\tt cor}({\vec x}_i,\lambda).
\label{eq:Qcorona}
\end{align}

Integrating the corona over the entire Einstein ring yields the spectral density shown in Fig.~\ref{fig:Qcorpix-c} (right).

\section{Sensitivity at various wavelengths}
\label{sec:SNR-all}

Fundamentally, there are two ways to block unwanted light from the Sun:
\begin{inparaenum}[i)]
\item An internal coronagraph uses an occulter that is built into the telescope and as such, is subject to its diffraction limit;
\item An external occulter, also known as a starshade, is located at a substantial distance from the telescope, producing an artifical eclipse.
\end{inparaenum}

\subsection{Modeling the internal coronagraph}
\label{sec:sol-coronagraph-model}

\begin{figure}
\includegraphics[]{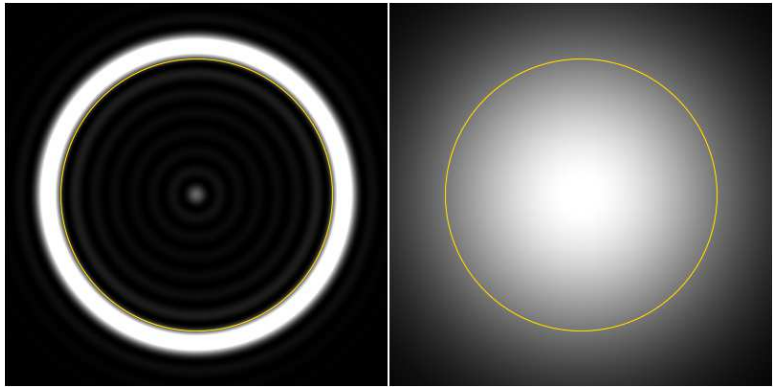}
\caption{\label{fig:ER-montage}Views by a $d=1$~m telescope at 650~AU, with the Sun blocked out (the size of the solar disk is indicated by the yellow circle.) Left: the Einstein ring of a point source, at $\lambda=1~\mu$m. Right: The completely blurred Einstein ring of the same point source, viewed through the same telescope at $\lambda=10~\mu$m.}
\end{figure}

In the case of an internal coronagraph, diffraction plays a critical role. An internal coronagraph uses an occulter mask in the focal plane of its first lens. Such an occulter is subject to the diffraction limit of the telescope aperture, therefore there will be significant diffracted light. To prevent this light from forming a bright Arago spot, a second lens is used that effectively transforms back the wavefront into the Fourier domain, where a smaller aperture, the Lyot-stop, filters out this diffracted light by acting as a low-pass filter. Finally, additional optics refocuses the light onto the telescope's image plane. The complexity of the optics and the inefficiency of the Lyot-stop reduce the transmittance of a coronagraph-equipped telescope to values as low as 12\% \cite{Zhou:2018,Turyshev-etal:2020-PhaseII}, which may need to be compensated by some combination of longer integration times or larger apertures.

For the internal coronagraph to be effective, the separation between the solar disk and the Einstein ring must be consistent with the telescope's diffraction limit. This pushes the beginning of the science operations for a meter-class telescope to heliocentric distances beyond $\bar z =650$~AU (see details in \cite{Turyshev-Toth:2020-image,Turyshev-Toth:2020-extend,Toth-Turyshev:2020,Turyshev-Toth:2022-wil_SNR}). At those distances, a $d=1$~m aperture telescope can resolve the Einstein ring and distinguish it from the Sun at the wavelength $\lambda=1~\mu$m; however, the same telescope at the same heliocentric ranges only sees an unresolved blur at $\lambda=10~\mu$m, as shown in Fig.~\ref{fig:ER-montage}~(right).

More specifically, resolving the Einstein ring from the solar disk implies the following condition:
{}
\begin{eqnarray}
\sqrt{\frac{2r_g}{\overline z}}-\frac{R_\odot}{\overline z}  \gtrsim \frac{\lambda}{d},
\label{eq:cor-gen}
\end{eqnarray}
in fact, $\sim2\lambda/d$ is needed for reliable operations.
Evidently, larger apertures are essential to operate at longer wavelengths. For a given wavelength $\lambda$, the size of the optimal aperture can be estimated from (\ref{eq:cor-gen}) as
{}
\begin{eqnarray}
d\gtrsim \lambda  \frac{\overline z}{\sqrt{2r_g \overline z}-R_\odot},
\label{eq:cor-gen_g}
\end{eqnarray}
which suggest that for $\lambda \simeq 1\,\mu$m, ideal apertures begin at $d\gtrsim 1.5$\,m. For $\lambda \simeq 10\,\mu$m, this optimal aperture increases to $d\simeq 15$\,m. As delivering such a large telescope to the SGL focal region is a complex engineering challenge, this limitation makes the use of an internal coronagraph for SGL observations at mid-IR wavelengths less desireable.

\subsection{The case for an external occulter (starshade)}

\begin{figure}
\includegraphics[scale=0.8]{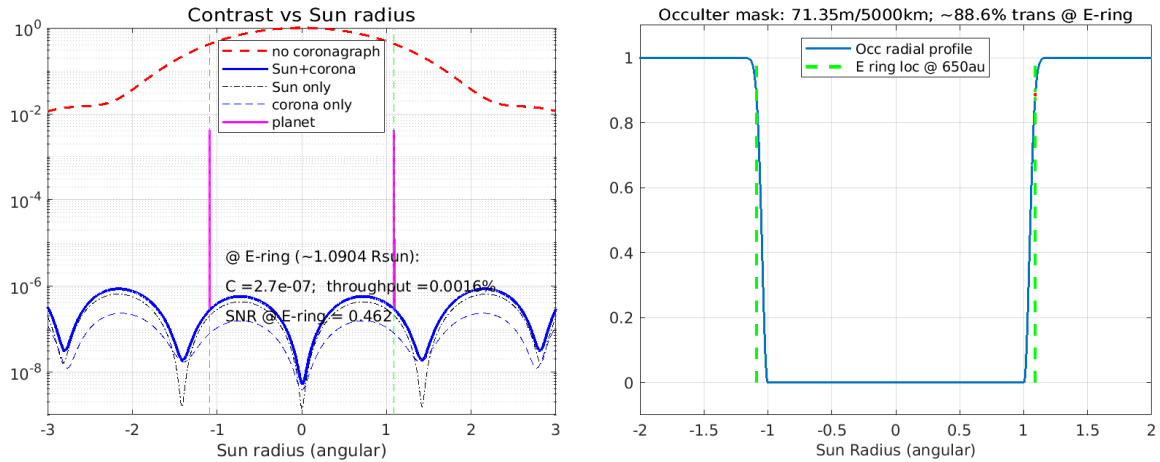}
\caption{\label{fig:starshade}Simulation of an external occulter (starshade) with a soft edge. The modeled Gaussian corrugation is $\sim$10\% of the occulter radius, which is very conservative. A smaller corrugation, which does not decrease transmittance at the Einstein ring, may be applicable. From \cite{Zhou:2022}.}
\end{figure}

For an external occulter, we need to have a physical obscuration that blocks the solar disk as seen from all points over the imaging telescope aperture, namely
{}
\begin{eqnarray}
\frac{D_0-d}{z_s}\gtrsim
\frac{2R_\odot}{\overline z},
\label{eq:cor-gen-ext}
\end{eqnarray}
where
$D_0$ is the diameter of the external occulter, $d$ is the telescope's aperture, and $z_s$ is the separation between the occulter and the telescope.

In comparison with projects that involve the use of a starshade for exoplanet detection in conjunction with a conventional space telescope \cite{Traub-Oppenheimer:2010,Cash:2011}, the starshade requirements of an SGL observatory are unremarkable. The required contrast ratio, $\sim 10^{7}$ at optical wavelengths, decreasing to as low as $10^{3}$ in mid-IR, is easily achievable. The purpose of the starshade is to block light from the Sun, as opposed to light from the host star. The angular separation between the Einstein ring and the solar disk is of ${\cal O}(0.1'')$. These requirements are easily satisfied by a starshade of modest size at a distance from the telescope that is about an order of magnitude smaller than the currently contemplated starshade-to-telescope distances; e.g., a 70~m starshade located at $\sim 5000$~km from an SGL telescope could block the Sun in a configuration with a Fresnel number of $F=(\tfrac{1}{2}D_0)^2/(\lambda z_s)\sim 25$ at $\lambda=10~\mu$m, sufficient for the efficient rejection of sunlight. Such a starshade is comparable in size to the solar sails that have been contemplated as the primary means of propulsion to deliver a telescope to the SGL focal region in acceptable timeframes. This raises the possibility that perhaps an appropriately designed, maneuverable solar sail package might be reused as a starshade during the science phase of the SGL mission. Recent simulations \cite{Zhou:2022} confirm the possibility of such a coronagraphic performance of a starshade (see Fig.~\ref{fig:starshade}).

The use of a starshade has other advantages, arising from the fact that the starshade is not subject to the diffraction limit imposed by the observing spacecraft's aperture.
\begin{itemize}
\item First, is not necessary to wait until ${\overline z}>650$~AU to begin observations; although the corona is brighter when the Einstein ring is closer to the Sun, the starshade reliably blocks light from the solar disk shortly after the spacecraft passes through ${\overline z}=548$~AU.
\item Second, the starshade can be positioned such that it blocks all light (including light from the corona) right up until the inner boundary of the Einstein ring, reducing the amount of corona noise.
\item Third, as we have known all along, as the spacecraft egresses, the apparent size of the Sun would change, and this would have to be mimicked in some way by the occulter.  With the starshade, we can simply vary the distance between starshade and spacecraft so that it always covers everything inside that Einstein ring.
\item Fourth, the modern starshade concepts that were studied by the exoplanetary community must be able to accommodate a significant slewing needed to repoint the pair starshade-observing telescope to observe a new target, making such concepts technically challenging. In contrast, in the case of the SGL, any motion would amount only to a pair of bore-sighted instruments (i.e., starshade and imaging telescope) slewing in the image plane, which is a substantially less demanding navigational task.
\item Fifth, an internal occulter requires complicated optics reducing optical throughput significantly, to as little as 12\% \cite{Zhou:2018,Turyshev-etal:2020-PhaseII}. A starshade may block sunlight without affecting the signal from the Einstein ring.
\item Lastly, we can conceive of an advanced starshade design with an occulter that has only an annular opening, allowing in light only from the very thin region of the Einstein ring (again keeping in mind that the starshade is not constrained by the diffraction limit of the observing telescope, only by the limits imposed by its ``soft'' edge), thus not only eliminating light contamination by the Sun but also drastically reducing light contamination from the solar corona.
\end{itemize}

Once an external starshade is used, the requirements on the SGL telescope can be greatly relaxed. The telescope is no longer required to resolve the Einstein ring from the solar disk in order for an internal coronagraph to block out the Sun. Furthermore, at longer wavelengths shot noise becomes a proportionately less significant problem, as the number of photons associated with a given signal power increases linearly with wavelength.

Using longer wavelengths in combination with a small telescope implies that no resolved Einstein ring can be seen anymore: any light that gets past the starshade is blurred completely by the telescope's diffraction limit, as shown in Fig.~\ref{fig:ER-montage}~(right). While its spectral resolution remains unaffected, the telescope can only be used in ``light bucket'' mode, moving from pixel location to pixel location in the image plane, measuring the intensity of received light for later reconstruction of a spatially resolved image of the target.

\subsection{Sensitivity at optical and near-IR wavelengths}
\label{sec:SNR}

At this point we are ready to evaluate the sensitivity that may be anticipated for different observing regimes. As always, we evaluate the sensitivity by computing the SNR. To do so, we assume that the unwanted light contamination from the solar corona can be entirely removed (e.g., by measuring the solar corona from a different vantage point, observing the same corona at the same time, with the target Einstein ring absent.) Even in this case, however, there is non-removable stochastic shot noise that is calculated as the square root of the total number of photons (signal plus light contamination) received. Therefore, we estimate the SGL SNR as
\begin{align}
{\tt SNR}(\vec{x}_0,\vec{x}_i,\lambda)\simeq
\frac{Q(\vec{x}_0,\vec{x}_i,\lambda)}{\sqrt{Q_{\tt cor}(\vec{x}_0,\vec{x}_i,\lambda)}},
\label{eq:theSNR}
\end{align}
where the approximation remains valid so long as $Q\gg Q_{\tt cor}$, which is almost always the case.

\begin{figure}
\includegraphics[]{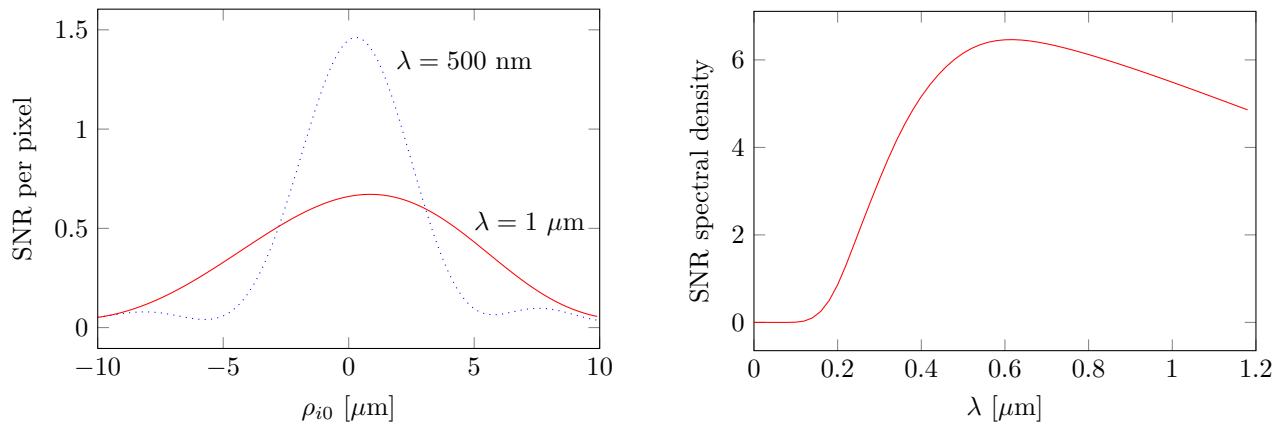}
\caption{\label{fig:SNRpix-c}Left: The per-pixel, per-spectral-channel SNR, for an Earth-like exoplanet at $z_0=10$~pc, as seen through a $d=1$~m aperture telescope with $f=10$~m focal distance, situated at ${\overline z}=650$~AU from the Sun, through the solar corona, after 300~s of integration time. Two cases are shown for $\lambda=1~\mu$m (solid red line) and $\lambda=500$~nm (dotted blue line) normalized to a nominal channel bandwidth of 1~$\mu$m.
Right: The SNR integrated over the Einstein ring in the visible part of the spectrum. A resolved Einstein ring is assumed, hence the bandwidth is limited to near-IR, $\lambda<1.2~\mu$m.
}
\end{figure}

To use Eq.~(\ref{eq:theSNR}), it is technically necessary to independently integrate the quantities in the numerator and under the square root in the denominator, over the integration time $t$, the spectral channel bandwidth $\lambda$, and the sensor (pixel) area $A$:

\begin{align}
{\tt SNR}\big({\vec x}_0,\Delta t, \Delta \lambda,A_{\tt }\big)= \ddfrac{ \int_{t_1}^{t_2} dt \int_{\lambda_1}^{\lambda_2} d\lambda \iint_A d^2{\vec x}_i~Q(\vec{x}_0,\vec{x}_i,\lambda)}{\Big[\int_{t_1}^{t_2} dt \int_{\lambda_1}^{\lambda_2} d\lambda \iint_A d^2{\vec x}_i~Q_{\tt cor}(\vec{x}_0,\vec{x}_i,\lambda)\Big]^\frac{1}{2}}.
\end{align}

This integration can only be carried out numerically. However, when the photon flux is constant over the integration time $\Delta t=t_2-t_1$, the sensor pixel size $A_{\tt pix}\simeq \Delta x_i \Delta y_i$, centered on $\vec{x}_i$, is small, and the spectral channel bandwidth $\Delta\lambda=\lambda_2-\lambda_1$ centered on $\lambda$ is narrow, the integration can be well approximated by simple multiplication:
\begin{align}
{\tt SNR}\big({\vec x}_0,\Delta t, \Delta \lambda,A_{\tt }\big)={\tt SNR}(\vec x_0,\vec x_{i}, t,\lambda)\sqrt{A_{\tt pix}~\Delta\lambda~\Delta t}.
\end{align}
It is, of course, also possible to just integrate over a range of wavelengths (to obtain a broadband per-pixel SNR for individual locations in the telescope sensor plane) or over the Einstein ring (or parts of the Einstein ring) to obtain a spectrally resolved SNR for the entire telescope sensor or for segments of the Einstein ring.

We illustrate these concepts first in Fig.~\ref{fig:SNRpix-c} (left), which shows (\ref{eq:theSNR}). To calculate the SNR, we assumed a per-pixel integration time of $\Delta t=300$~seconds: that is to say, the telescope is expected to collect light continuously at a fixed location relative to the (moving) exoplanet image for 300 seconds before moving on to the next location. This integration time is sufficiently short to avoid motion blur, e.g., due to planetary rotation. (Other effects due to temporal changes in the planet's phase \cite{Traub-Oppenheimer:2010} or illumination will be considered in future studies.) For context, consider that the area of the $\lambda=1~\mu$m Einstein ring occupies approximately 4,900~$\mu$m$^2$ in the sensor plane of this telescope. This result is consistent with our earlier estimates \cite{Turyshev-Toth:2020-im-extend,Toth-Turyshev:2020}, demonstrating the feasibility of obtaining intermediate-to-high resolution images of exoplanets in the 5--30~pc range using the SGL and a $d=1$~m telescope.

Further integrating over the entire Einstein-ring yields Fig.~\ref{fig:SNRpix-c} (right). This figures depicts the SNR at various optical and near-IR wavelengths, assuming that the optical telescope was used as a single-pixel, spectrally resolved sensor sampling the SGL image plane at $\vec{x}_0$.

\begin{figure}
\includegraphics[]{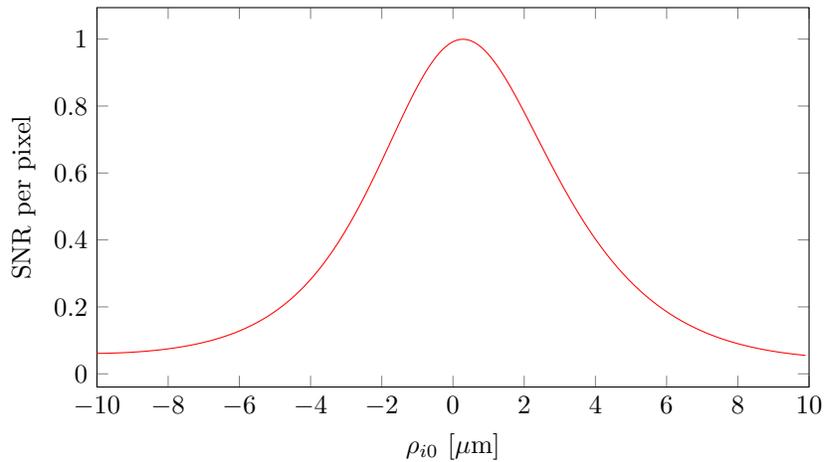}
\caption{\label{fig:SNRl}Broadband SNR in the visible and near-IR band, $200~{\rm nm}\le\lambda\le 1.2~\mu$m, shown over the cross-section of a resolved Einstein ring, over a 300-second integration time using a $d=1$~m telescope, consistent with Fig.~\ref{fig:SNRpix-c}.}
\end{figure}

Conversely, Fig.~\ref{fig:SNRl} shows the SNR per pixel after integrating over all optical and near IR wavelengths up to $\lambda=1.2~\mu$m. This would correspond to an observational scenario that uses details of the Einstein ring to help reconstruct broadband, spatially resolved high-resolution images of the source.

Other combinations are, of course, possible, including integration over both the entire Einstein ring and over all wavelengths. This scenario, using a $d=1$~m telescope with $f=10$~m focal line, imaging the Einstein ring of an exo-Earth at $z_0=10$~pc, from a vantage point at ${\overline z}=650$~AU, yields the value of ${\tt SNR}=12.67$ after 300 seconds of integration.

\begin{figure}
\includegraphics[]{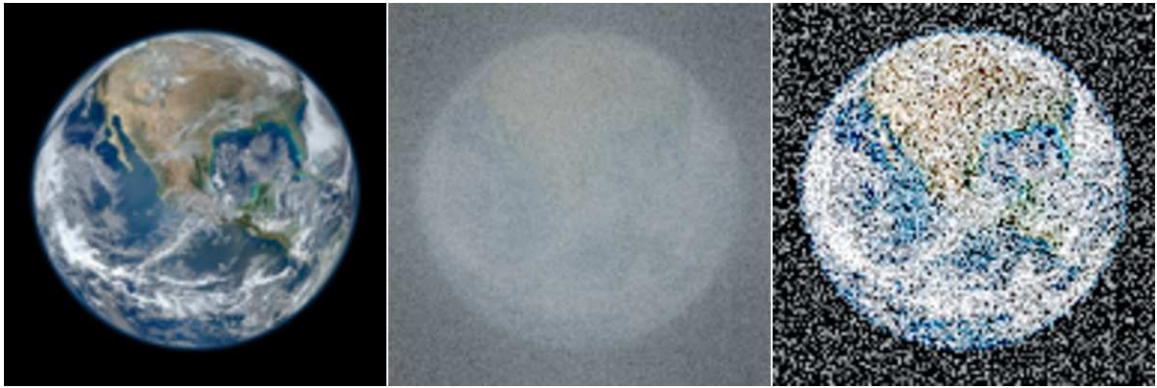}
\caption{\label{fig:earthC128}RGB color observation of an exo-Earth at 30~pc using the SGL in conjunction with a $d=1$~m telescope equipped with an internal coronagraph. Left: the source image at $128\times 128$ pixel resolution; Center: the blurred image projected by the SGL into an image plane at ${\overline z}=650~$AU, with nonremovable stochastic shot noise from the corona at ${\tt SNR}=12.67$, consistent with our solar corona model at 650~AU with 300-second per pixel, per color channel integration time, or a cumulative integration time of 6 months. Right: image after deconvolution, at ${\tt SNR}=1.05$ per channel.
}
\end{figure}

This SNR is sufficient to reconstruct even a color image of an exo-Earth at a modest resolution. This is shown in Fig~\ref{fig:earthC128}, which shows the result consistent with a cumulative total integration time of less than 6 months. The post-deconvolution per-channel SNR of $\sim 1.05$ is consistent with the semianalytically estimated SNR \cite{Turyshev-Toth:2022-wil_SNR} of
\begin{align}
{\tt SNR}_{\tt R}\simeq 0.891\frac{D}{d{\sqrt N}}{\tt SNR}_{\tt C}=0.90,
\label{eq:SNRR}
\end{align}
where $D=1300~{\rm m}/128=10.15$~m is the spacing between pixels.

\subsection{Observations at longer wavelengths}
\label{sec:IR}

In the preceding section, we evaluated the use of the SGL in combination with a meter-class aperture, assuming a resolved Einstein ring. This scenario allows the use of an internal coronagraph or other observational techniques that rely on sampling parts of the Einstein ring. We observed that the diffraction limit of such a small telescope confines us to optical and near IR wavelengths. In particular, at longer wavelengths, the telescope's diffraction limit stands in the way (see Fig.~\ref{fig:ER-montage}) of resolving the solar disk apart from the Einstein ring that surrounds it, making it impossible to block light contamination by the Sun using an internal coronagraph.

Yet as we noted earlier, an Earth-like exoplanet emits significant power at longer wavelengths, in the mid-IR, thermal regime (see Fig.~\ref{fig:Qs}). Moreover, the solar corona is significantly quieter at these wavelengths (see Fig.~\ref{fig:Qcor}), which should result in a much improved SNR. To sidestep the need for a telescope with an extremely large aperture, we consider another possibility: An external starshade (or, specifically, sunshade).

The spectral sensitivity of the SGL in the mid-IR domain can be calculated as before, using (\ref{eq:pow-cor*}). (One intriguing possibility that may be within the realm of technical feasibility is to extend the starshade to also block light from {\em outside} the Einstein-ring, effectively implementing an ``annular coronagraph'' concept that we first discussed in \cite{Turyshev-Toth:2020-extend}. This approach would amount to reducing the integration limits in (\ref{eq:pow-cor*}) from $\theta_0..\infty$ to a narrow range centered on $\theta_{\tt ER}=\sqrt{{2r_g}/{\overline z}}$, characterizing the annular opening in this starshade concept. The width of the annular opening is no longer constrained by the imaging telescope's diffraction limit either, only by the size of the starshade itself and the nature of its patterned ``soft'' edges, which are used to prevent the appearance of the bright Arago spot that would otherwise be formed by such a disk-shaped occulter.)

\begin{figure}
\includegraphics[]{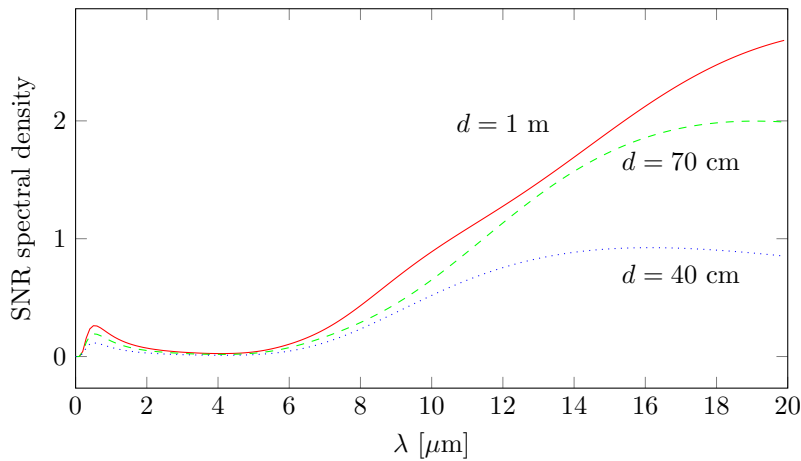}
\caption{\label{fig:SNRIR-l}The use of an external starshade allows us to extend observations beyond that shown in Fig.~\ref{fig:SNRpix-c}, into the mid-IR domain. In this domain, the much higher photon count and comparatively less noise from the corona, offers remarkable capabilities, including the use of smaller telescopes. Depicted is the broadband spectral SNR density, using only a 1~s integration time and integrated over the telescope sensor, with an assumed field-of view twice the diameter of the apparent size of the Sun, using three different aperture sizes: $d=1$~m (red solid line), $d=70~$cm (green dashed line) and $d=40~$cm (blue dotted line). Note the wide applicable bandwidth with wavelengths ranging from UV (100~nm) to mid-IR (20~$\mu$m).}
\end{figure}

It is interesting that even a simple disk starshade makes the use of much smaller telescopes feasible even in the mid-IR band. Fig.~\ref{fig:SNRIR-l} demonstrates that even a modest 40~cm can deliver an SNR that is sufficient for spatially resolved spectroscopy in a range of bandwidths that is of great interest to the exobiology community.

\begin{table}
\caption{\label{tb:SNRIR}The broadband SNR (i.e., for the bandwidth within $0.1~\mu{\rm m}\le\lambda\le 20~\mu{\rm m}$)  for a range of telescope apertures $d$, while using a starshade to image an exo-Earth at $z_0=$~30~pc from the SGL focal region at ${\overline z}=$~650~AU.}
\begin{tabular}{|r|c|}\hline
~~~~~$d$~~~~~&~~${\tt SNR}$ in 1 sec~~\\\hline\hline
0.4~m&0.95\\
0.7~m&2.18\\
1.0~m&3.30\\
\hline
\end{tabular}
\end{table}

The broadband SNR (i.e., for the bandwidth within $0.1~\mu{\rm m}\le\lambda\le 20~\mu{\rm m}$) for a range of telescope apertures looking at an exo-Earth at $z_0=$~30~pc from an observing location at ${\overline z}=650$~AU, is shown in Table~\ref{tb:SNRIR}. To demonstrate the potential utility of these observations, we considered a 40~cm telescope in this configuration. We again assumed a per-pixel integration time of 300~s, which, according to Table~\ref{tb:SNRIR}, translates into an effective SNR of $0.95\sqrt{300}\sim 16.5$.

\begin{figure}
\includegraphics[]{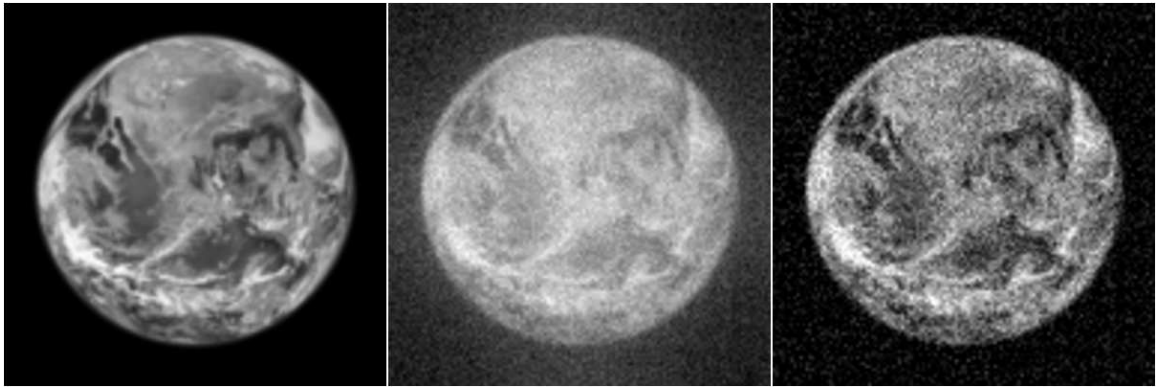}
\caption{\label{fig:earth128}Observing an exo-Earth at 30~pc via the SGL, using a small, 40-cm telescope in the mid-, or thermal infrared part of the spectrum (broadband signal with $\lambda\le 20~\mu$m) in conjunction with an external starshade/sunshade that is assumed to block out the Sun. Left: the source image at $128\times 128$ pixel resolution; Center: the blurred image projected by the SGL into an image plane at ${\overline z}=650~$AU, with nonremovable stochastic shot noise from the corona at ${\tt SNR}=16.5$, consistent with our solar corona model at 300-second per pixel integration time, or a cumulative integration time of 2~months. Right: image after deconvolution. The use of a starshade makes such a remarkably small instrument (i.e., with aperture of only 40 cm) capable of delivering high-quality spatially or spectrally resolved images across a broad range of wavelengths.}
\end{figure}

Fig.~\ref{fig:earth128} shows the results of this simulation at an image resolution of $N=128\times 128$ image pixel locations. The original image (using an image of the actual Earth as a stand-in for the exoplanet) is, as expected, blurred by the SGL; the middle panel shows the blurred image with Gaussian noise added at the level of ${\tt SNR}_{\tt C}=16.5$. Deconvolution yields the image on the right, with ${\tt SNR}_{\tt R}=3.04$, again consistent with (\ref{eq:SNRR}).
Note that an observation of $128\times 128=16,384$ pixels with 300 seconds per pixel can be accomplished using less than two months of cumulative integration time.

It is, of course, possible to trade spatial resolution for spectral resolution. Reduced spatial resolution also allows for an increase in integration times without concern for motion blur. Therefore, we anticipate that even with a modest, $d=40~$cm telescope, spatially resolved spectroscopy of a distant exoplanet with the number of spectral channels in excess of 100, covering wavelengths from UV to mid-IR, may be possible if an external starshade is used.

\section{Conclusions and next steps}
\label{sec:concl}

One of the more significant challenges to using the solar gravitational lens to image distant, faint targets is the fact that any light from such targets appears in the form of an Einstein ring on top of the bright solar corona. For a target such as an Earth-like exoplanet, despite the significant light amplification of the SGL, the solar corona remains brighter than the target's Einstein ring by several orders of magnitude.

As the solar corona background is known (in fact, it can be independently measured, e.g., by nearby spacecraft that ``sees'' the same corona but not the Einstein ring of the intended target) it can be removed. Needless to say, this implies an observing instrument that has the requisite dynamic range to detect minute variations of the faint Einstein ring on top of that bright background. More importantly, due to the quantized nature of light, removing the corona background still leaves a non-removable stochastic noise component: the Poisson ``shot'' noise.

When a large number of photons are involved, Poisson noise is indistinguishable from Gaussian noise. As such, its contribution can be readily estimated dividing the number of signal photons by the square root of total photons. Previously, in \cite{Turyshev-Toth:2020-extend}, we estimated the broadband SNR for the entire Einstein ring of an Earth-like exoplanet.

In the present paper, we extended this work by rigorously developing estimates of photon count spectral densities at individual pixel locations in the focal plane of an imaging telescope. By doing so, we have now developed the tools that are needed to understand key questions, such as the achievable spectral resolution of a future SGL instrument, or the extent to which details of the Einstein ring may be used to aid or improve image reconstruction.

Our calculations are based on a previously developed wave-optical treatment of the SGL. To develop the present results, we treated the Sun as a gravitational monopole. We ignored contributions from its quadrupole and higher mass multipole moments. These contributions are important when it comes to image reconstruction, but they have no impact on the observed brightness of the solar corona, and negligible impact on the overall brightness of the observed Einstein ring of an extended body so long as its projected image is much larger than the quadrupole-induced caustic pattern in the SGL image plane \cite{Turyshev-Toth:2021-multipoles,Turyshev-Toth:2021-caustics,Turyshev-Toth:2021-quartic}. This condition is easily satisfied in all the realistic cases that we have considered (i.e., an Earth-like planet with habitable zone and within 30~pc from us).

A future SGL mission is expected to bring images with good spatial and spectral resolution. The results we presented in this paper allow us to estimate the anticipated SNR per spectral channel. This information is useful for instrument and mission design; it will also be valuable to evaluate, inform, and guide prospective science observations.

We also considered using the SGL in the mid-IR band. This was not previously considered, as a telescope with an internal coronagraph would have to be unreasonably large to resolve the solar disk from the Einstein ring at wavelengths up to $\lambda=20~\mu$m. In this case, however, we considered the use of an external occulter or sunshade, similar to the starshades that are currently being considered for exoplanet search campaigns \cite{Traub-Oppenheimer:2010}. We found that the starshade requirements of the SGL are more modest than these proposed starshades. As the size of the required starshade is comparable to that of the solar sail that is being contemplated as the primary propulsion method for an SGL mission \cite{Turyshev-etal:2020-PhaseII}, this opens up the possibility that the solar sail may also double as a starshade.

Once a starshade is used, a much smaller SGL telescope can deliver useful results across a broad range of wavelengths from UV to optical to mid-IR and beyond. The resulting SNR is quite remarkable, as terrestrial planets are strong signal emitters in the mid-IR range of wavelengths, where corona noise is comparatively less. Our simulations show that, quite remarkably, even a 40~cm aperture telescope is sufficient to recover a good quality, resolved image of an exo-Earth as far as 30~pc from the Earth. These results offer strong motivation to study the use navigable starshades (and especially, solar sails repurposed as starshades) as essential components of future SGL missions.

The mission implications of using starshades (single or multiple) in conjunction with the SGL (that by itself is a subject of various dynamical motions, see discussion in \cite{Turyshev-Toth:2022-wobbles}) is a novel topic that would have to be investigated. In particular, specific imaging strategies for prospective exoplanet targets are yet to be worked out, taking into account the target system's dynamics and also the temporal behavior of the target system and planet. This effort is on-going and results, when available, will be reported elsewhere.

\begin{acknowledgments}
The authors thank Michael Shao and Hanying Zhou at JPL for extremely valuable discussions.
This work in part was performed at the Jet Propulsion Laboratory, California Institute of Technology, under a contract with the National Aeronautics and Space Administration.
VTT acknowledges the generous support of Plamen Vasilev and other Patreon patrons.
\end{acknowledgments}

\pagebreak[2]


\end{document}